\pdfoutput=1

\documentclass[journal]{IEEEtran}
\ifCLASSINFOpdf
  \usepackage[pdftex]{graphicx}
   \graphicspath{{../pdf/}{../jpeg/}}
   \DeclareGraphicsExtensions{.pdf,.jpeg,.png}
\else
   \usepackage[dvips]{graphicx}
   \graphicspath{{../eps/}}
   \DeclareGraphicsExtensions{.eps}
\fi
\ifCLASSOPTIONcompsoc
\usepackage[caption=false,font=normalsize,labelfon
t=sf,textfont=sf]{subfig}
\else
\usepackage[caption=false,font=footnotesize]{subfi
	g}
\fi

\usepackage{amsmath}
\usepackage{algorithm}  
\usepackage{algorithmic}
\begin{document}
%
% paper title
% Titles are generally capitalized except for words such as a, an, and, as,
% at, but, by, for, in, nor, of, on, or, the, to and up, which are usually
% not capitalized unless they are the first or last word of the title.
% Linebreaks \\ can be used within to get better formatting as desired.
% Do not put math or special symbols in the title.
\title{Modifications of \emph{FastICA} in Convolutive Blind Source Separation  }
%
%
% author names and IEEE memberships
% note positions of commas and nonbreaking spaces ( ~ ) LaTeX will not break
% a structure at a ~ so this keeps an author's name from being broken across
% two lines.
% use \thanks{} to gain access to the first footnote area
% a separate \thanks must be used for each paragraph as LaTeX2e's \thanks
% was not built to handle multiple paragraphs
%

\author{YunPeng ~Li
         
         % <-this % stops a space
\thanks{YunPeng Li was with the Department of Automation, Tsinghua University, Beijing, China e-mail: liyp18@mails.tsinghua.edu.cn}
}

% note the % following the last \IEEEmembership and also \thanks - 
% these prevent an unwanted space from occurring between the last author name
% and the end of the author line. i.e., if you had this:
% 
% \author{....lastname \thanks{...} \thanks{...} }
%                     ^------------^------------^----Do not want these spaces!
%
% a space would be appended to the last name and could cause every name on that
% line to be shifted left slightly. This is one of those "LaTeX things". For
% instance, "\textbf{A} \textbf{B}" will typeset as "A B" not "AB". To get
% "AB" then you have to do: "\textbf{A}\textbf{B}"
% \thanks is no different in this regard, so shield the last } of each \thanks
% that ends a line with a % and do not let a space in before the next \thanks.
% Spaces after \IEEEmembership other than the last one are OK (and needed) as
% you are supposed to have spaces between the names. For what it is worth,
% this is a minor point as most people would not even notice if the said evil
% space somehow managed to creep in.

% The paper headers
\markboth{MANUSCRIPT}%
{Shell \MakeLowercase{\textit{et al.}}: Bare Demo of IEEEtran.cls for IEEE Journals}
% The only time the second header will appear is for the odd numbered pages
% after the title page when using the twoside option.
% 
% *** Note that you probably will NOT want to include the author's ***
% *** name in the headers of peer review papers.                   ***
% You can use \ifCLASSOPTIONpeerreview for conditional compilation here if
% you desire.

% If you want to put a publisher's ID mark on the page you can do it like
% this:
%\IEEEpubid{0000--0000/00\$00.00~\copyright~2015 IEEE}
% Remember, if you use this you must call \IEEEpubidadjcol in the second
% column for its text to clear the IEEEpubid mark.

% use for special paper notices
%\IEEEspecialpapernotice{(Invited Paper)}

% make the title area
\maketitle

% As a general rule, do not put math, special symbols or citations
% in the abstract or keywords.
\begin{abstract}
Convolutive blind source separation (BSS) is intended to recover the unknown components from their convolutive mixtures. Contrary to the  contrast functions used in instantaneous cases, the  spatial-temporal prewhitening stage and the para-unitary filters constraint are difficult to implement in a convolutive context. In this paper, we propose several modifications of \emph{FastICA} to alleviate these difficulties. Our method performs the simple prewhitening step on convolutive mixtures prior to the separation and optimizes the contrast function under the diagonalization constraint implemented by single value decomposition (SVD). Numerical simulations are implemented to verify the performance of the proposed method.
\end{abstract}

% Note that keywords are not normally used for peerreview papers.
\begin{IEEEkeywords}
Convolutive blind source separation (BSS), FastICA, diagonalization constraint, singular value decomposition (SVD). 
\end{IEEEkeywords}

% For peer review papers, you can put extra information on the cover
% page as needed:
% \ifCLASSOPTIONpeerreview
% \begin{center} \bfseries EDICS Category: 3-BBND \end{center}
% \fi
%
% For peerreview papers, this IEEEtran command inserts a page break and
% creates the second title. It will be ignored for other modes.
\IEEEpeerreviewmaketitle

\section{Introduction}
% The very first letter is a 2 line initial drop letter followed
% by the rest of the first word in caps.
% 
% form to use if the first word consists of a single letter:
% \IEEEPARstart{A}{demo} file is ....
% 
% form to use if you need the single drop letter followed by
% normal text (unknown if ever used by the IEEE):
% \IEEEPARstart{A}{}demo file is ....
% 
% Some journals put the first two words in caps:
% \IEEEPARstart{T}{his demo} file is ....
% 
% Here we have the typical use of a "T" for an initial drop letter
% and "HIS" in caps to complete the first word.
\IEEEPARstart{C}{onvolutive} blind source separation (BSS) has received much attention in recent years. It has been successfully applied in many signal processing problems, such as sonar array processing, seismic exploration, and the "cocktail party problem". In the convolutive BSS, $n$ dimensional observation $\mathbf{x}(k)=\left(x_{1}(k),\cdots,x_{n}(k) \right)^{T}$ is the convolutive mixtures of the $m$ dimensional independent source $\mathbf{s}(k)=\left(s_{1}(k),\cdots,s_{m}(k) \right)^{T}$. The unknown mixing process $\mathbf{A}$ can be described by a Multi-Input Multi-Output (MIMO) linear time invariant (LTI) system $\left( \mathbf{A}(k)_{k\in Z}\right)$,
\begin{equation}
\label{eq:mixing}
\mathbf{x}(k)=\mathbf{A}(k)*\mathbf{s}(k)=\sum_{l\in Z}\mathbf{A}(l)\mathbf{s}(k-l)
\end{equation}

Given $N$ samples of observation $\mathbf{x}(k)$, recovering the sources' estimation $\mathbf{y}(k)=\left(y_{1}(k),\cdots,y_{m}(k) \right)^{T}$ from these mixtures is equivalent to find a filter banks $\left( \mathbf{B}(k)_{k\in Z}\right)$ to
inverse the mixing system $\left( \mathbf{A}(k)_{k\in Z}\right)$:
\begin{equation}
\label{eq:unmixng}
\mathbf{y}(k)=\mathbf{B}(k)*\mathbf{x}(k)=\sum_{l\in Z}\mathbf{B}(l)\mathbf{x}(k-l)
\end{equation}
each $y_{i}(k)$ is a scaling and filtering version of the unique source's component $s_{i^{'}}(k)$.
\par Although the ordering and scaling ambiguity in the instantaneous BSS can be efficiently handled by the prewhitening on $\mathbf{x}(k)$ and the orthogonal constraint of unmixing matrix $\mathbf{B}$, the filtering ambiguity caused by the time delay is difficult to deal with in convolutive context and these strategies (used in instantaneous BSS) become invalid. 
These ambiguities lead to much more extreme points in convolutive BSS than in instantaneous BSS, making the convolutive BSS a tough problem to deal with.
\par In this paper, we shall assume that:
\begin{enumerate}[]
	\item The source signals $\mathbf{s}(k)$ are real-valued, zero-mean, and mutually statistically independent, at most one of them is Gaussian.
	\item The filter banks $\left( \mathbf{A}(k)_{k\in Z}\right)$, $\left( \mathbf{B}(k)_{k\in Z}\right)$ are stable, causal, and finite impulse response (FIR).
\end{enumerate}
\par Many Methods\cite{handbook2010}\cite{Pedersen2007A} have been proposed to solve the convolutive BSS. Most of them can be classified into two groups: the frequency domain approaches and the time domain approaches. In frequency domain\cite{dye1994}\cite{paris1998}, convolutive BSS can be considered as instantaneous BSS for each frequency bin, where each bin has own scaling and ordering indeterminacy as mentioned before. Complex value after discrete Fourier transform (DFT) and circularity problem happen in frequency domian.
\par Time domain approaches include density matching methods and contrast function methods. Density matching methods apply the well-known InfoMax\cite{infomax} proposed in ICA to the convolutive case\cite{Torkkola96}. The performance of the density matching approaches is highly dependent on
the prior knowledge on the unknown density distributions of $s_{i}$. It's important to determine whether the source $s_{i}(k)$ is super-Gaussian or 
sub-Gaussian beforehand. Density matching methods linearly transform the observed mixtures $\mathbf{x}(k)$ with the demixing system  $\left( \mathbf{B}(k)_{k\in Z}\right)$, forcing the 
$p_{y}(y_{i})$ close to a selected density $p_{s}(s_{i^{'}})$. Most of these methods are based on the gradient optimization, requiring appropriate choice of learning rate and step direction. Natural gradient\cite{nagrd1998} and relative gradient\cite{regrd1998} have been proposed to alleviate the drawbacks of stability and convergence above. Contrast function methods make the advantage of the statistical independence or non-Gaussianity of the components of source to recover the unknown $s_{i}$. In addition, High order statistics (HOS)\cite{ctr1996} of source's estimation $y_{i}(k)$ like cumulants, cross-cumulants, and cross-moments can work as contrast function in separation. Prewhitening stage and coefficents constraints are required to guarantee the uniqueness of the extracted components during the contrast function methods. For instantaneous case, prewhitening stages and orthogonal constraint can be implemented by Gram–Schmidt orthogonalization or singular value decomposition (SVD). While in the convolutive mixtures, the corresponding spatial-temporal prewhitening stage and
the para-unitary filters constraint are more difficult to realize.
\par \emph{FastICA}\cite{AHy1999} is a contrast function method in instantaneous BSS by maximizing the approximation of negentropy\cite{AHy1998}. It works well in instantaneous case for its fast convergence and robustness. Unfortunately, it cannot directly be adapted to the convolutive BSS due to the filtering ambiguity caused by the time delay. Several convolutive extensions of \emph{FastICA} have been proposed\cite{thm2006}\cite{scd2007} in recent years. The method proposed in \cite{thm2006} conducts a convolutive prewhitening to the transformed observation $\mathbf{x}(k)$ at first, then removes the extracted estimations $y_{i}$ at previous steps from the mixtures $\mathbf{x}$ in a deflation mode, leading to an accumulation of estimation errors which may become excessive after a certain number of source extractions. In \cite{scd2007}, a spatial-temporal prewhitening stage and the para-unitary constraint is implemented at the cost of high computation burden. The existing \emph{FastICA} extensions in convolutive BSS are difficult and inefficient.

In this paper, we propose a novel extension of \emph{FastICA}, under a simpler framework. The proposed modifications combine a convolutive prewhitening stage for observations with the diagonalization constraint in both deflation or symmetric mode.
\par The rest of the paper is organized as follows. In Section \ref{sec:2}, we state several assumptions of our method and rearrange the convolutive mixtures into instantaneous mixtures. An optimization problem based on the \emph{FastICA} is described in Section \ref{sec:3}.
We describe the prewhitening strategy along with the diagonalization constraint in Section \ref{sec:4}. Experiments' results are presented to verify our modifications in Section \ref{sec:5}. We conclude our method in Section \ref{sec:6}.

\section{Problem statement}
\label{sec:2}
We describe the mixing system $\mathbf{A}$ by the following FIR filter equation,
\begin{equation}
\label{x}
x_{i}(k) =\sum_{j=1}^{m}\sum_{l=0}^{P-1}a_{ij}(l)s_{j}(k-l) \quad \quad i=1,\cdots,n
\end{equation}
where the $\mathbf{a}_{ij}$ are the mixing filters. Without loss of generality, we assume all the mixing filters have the same filter order $P$.
The estimation of the demixing system $\mathbf{B}$ shares the same form,
\begin{equation}
\label{y}
y_{i}(k) =\sum_{j=1}^{n}\sum_{l=0}^{Q-1}b_{ij}(l)x_{j}(k-l) \quad \quad i=1,\cdots,m
\end{equation}
where the $\mathbf{b}_{ij}$ are the demixing filters and we assume all the demixing filters have the same filter order $Q$. $y_{i}(k)$ is a unique scaled, permuted, and filtered vesion of$s_{i^{'}}(k)$.
To reconstruct the contributions in each observation $x_{i}(k)$, another assumption is considered:
\begin{enumerate}
	\item Each signal source $s_{i}(k)$ is produced by an innovation process $u_{i}(k)$ via a stable FIR filters $F_{i}(k)$, where $u_{i}(k)$ is zero-mean, mutually independent and non-gaussian random process.
\end{enumerate}
Then, signal source $s_{i}(k)$ can be expressed in the following form,
\begin{equation}
\label{innovation}
s_{i}(k)=\sum_{l=-R+1}^{R-1}F_{i}(l)u_{i}(k-l) \quad \quad i=1,\cdots,m
\end{equation}
The order of the non causal  FIR $F_{i}$ is $2R-1$, all the  filter banks   $\mathbf{F}(k)=diag(F_{1}(k),\cdots,F_{m}(k))$ are connected with 
innovation process $\mathbf{u}(k)=\left(u_{1}(k),\cdots,u_{m}(k) \right)^{T}$ respectively. In order to individually extract $s_{i}$, we regard the $\mathbf{F}(k)$ as the coloring filters for the innovation process $\mathbf{u}(k)$.
\par To rearrange the convolutive BSS to instantaneous BSS, we created several variables as follow,
\begin{equation}
\label{s1}
\underline {\mathbf{s}}(k)=\left(\mathbf{s}_{1}^{T}(k),\mathbf{s}_{2}^{T}(k),\cdots,\mathbf{s}_{m}^{T}(k)\right)^{T}
\end{equation}

\begin{equation}
\label{s2}
\mathbf{s}_{i}(k) = \left(s_{i}(k),s_{i}(k-1),\cdots,s_{i}(k-P+1)\right)^{T}
\end{equation}
where $\underline{\mathbf{s}}(k)$ is  a $mP\times1$ column vector, and $\mathbf{s}_{i}(k)$ is a $P\times1$ column vecto. The mixing system $\mathbf{A}$ can be expressed in the matrix form,
\begin{equation}
\label{A1}
\mathbf{A}= \left(\begin{IEEEeqnarraybox*}[][c]{,c/c/c/c,}
\mathbf{a}_{11}^{T}&\mathbf{a}_{12}^{T}&\cdots &\mathbf{a}_{1m}^{T}\\
\mathbf{a}_{21}^{T}&\mathbf{a}_{22}^{T}&\cdots &\mathbf{a}_{2m}^{T}\\
\vdots&\vdots &\ddots &\vdots\\
\mathbf{a}_{n1}^{T}&\mathbf{a}_{n2}^{T}&\dots &\mathbf{a}_{nm}^{T}%
\end{IEEEeqnarraybox*}\right)
\end{equation}
\begin{equation}
\label{A2}
\mathbf{a}_{ij} = \left(a_{ij}(0),a_{ij}(1),\cdots,a_{ij}(P-1)\right)^{T}
\end{equation}
where $\mathbf{A}$ is a $n\times mP$ matrix, and $\mathbf{a}_{ij}$ is $P\times1$ column vector. The mixing system (\ref{x}) becomes, 
\begin{equation}
\label{mix1}
\mathbf{x}(k)=\mathbf{A}\underline{\mathbf{s}}(k)
\end{equation}
The demixing process can be transformed in the same way.

\begin{equation}
\label{x1}
\underline {\mathbf{x}}(k)=\left(\mathbf{x}_{1}^{T}(k),\mathbf{x}_{2}^{T}(k),\cdots,\mathbf{x}_{n}^{T}(k)\right)^{T}
\end{equation}

\begin{equation}
\label{x2}
\mathbf{x}_{i}(k) = \left(x_{i}(k),x_{i}(k-1),\cdots,x_{i}(k-Q+1)\right)^{T}
\end{equation}
where $\underline{\mathbf{x}}(k)$ is  a $nQ\times1$ column vector, and $\mathbf{x}_{i}(k)$ is a $Q\times1$ column vector.
\begin{equation}
\label{B1}
\mathbf{B}= \left(\begin{IEEEeqnarraybox*}[][c]{,c/c/c/c,}
\mathbf{b}_{11}^{T}&\mathbf{b}_{12}^{T}&\cdots &\mathbf{b}_{1n}^{T}\\
\mathbf{b}_{21}^{T}&\mathbf{b}_{22}^{T}&\cdots &\mathbf{b}_{2n}^{T}\\
\vdots&\vdots &\ddots &\vdots\\
\mathbf{b}_{m1}^{T}&\mathbf{b}_{m2}^{T}&\dots &\mathbf{b}_{mn}^{T}%
\end{IEEEeqnarraybox*}\right)
\end{equation}
\begin{equation}
\label{B2}
\mathbf{b}_{ij} = \left(b_{ij}(0),b_{ij}(1),\cdots,b_{ij}(Q-1)\right)^{T}
\end{equation}
where $\mathbf{B}$ is a $m\times nQ$ matrix, and $\mathbf{b}_{ij}$ is $Q\times1$ column vector.
\begin{equation}
\label{demix1}
\mathbf{y}(k)=\mathbf{B}\underline{\mathbf{x}}(k)
\end{equation}
\par Owing to the concise expressions of the mixing process (\ref{mix1}) and demixng process (\ref{demix1}), we have converted
the $m$ signals $n$ observations covolutive BSS into a $m$ signals $nQ$ observations instantaneous BSS problem. For the sake of uniqueness in extraction, a prewhitening stage is required in contrast function methods:
\begin{equation}
\label{prew}
\mathbf{v}(k) =\mathbf{H}(\mathbf{x}(k))
\end{equation}
we represent the prewhitening stage in function $\mathbf{H}()$, and the prewhitening ouput $\mathbf{v}(k)$ should comply with some constraint. In instantaneous case, the particular step can be conducted via eigenvalue decomposition or PCA,
\begin{equation}
\label{prew2}
\mathbf{v}(k) =\mathbf{H}\mathbf{x}(k)
\end{equation}
where $\mathbf{v}(k)$ conforms to the below constraint.
\begin{equation}
\label{prew3}
\boldmath{E}\{\mathbf{v}(k)\mathbf{v}^{T}(k)\}=\mathbf{I}_{n\times n}
\end{equation}
Unfortunately, the above prewhitening strategy fails in the convolutive case, more details are provided in Section \ref{sec:4}.
\begin{equation}
\label{sep}
\mathbf{y}(k)=\mathbf{B}\underline{\mathbf{x}}(k)=\mathbf{W}\mathbf{H}\underline{\mathbf{x}}(k)=\mathbf{W}\mathbf{v}(k)
\end{equation}
\par After prewhitening stage, our method (\ref{sep}) adjusts coefficients in separation matrix $\mathbf{W}$ to recover the $y_{i}(k)$ as an estimation of a delayed scaled innovation process $u_{i^{'}}(k-l)$ . 
\begin{figure}[!t]
	\centering
	\includegraphics[width=2.5in]{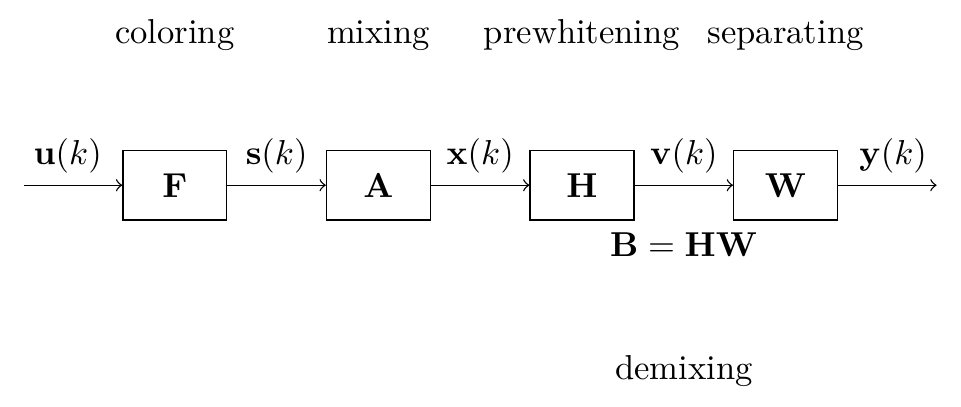}
	\caption{Flow chart of the convolutive BSS.}
	\label{flowchart}
\end{figure}
The complete routine of the convolutive BSS can be concluded in Fig.\ref{flowchart}. The source's component $\mathbf{s}(k)$ is the output of
the innovation process $\mathbf{u}(k)$ filtered by the coloring filters $\mathbf{F}$. Unknown mixing system $\mathbf{A}$ mixes the $\mathbf{s}(k)$ both in time and space. Given the observation
$\mathbf{x}(k)$, the proposed method conducts a demixing procedure $\mathbf{B}$ to recover $\mathbf{y}(k)$. The prewhitening $\mathbf{H}$ stages and seperating $\mathbf{W}$ stages are the keys in the demixing system.
\section{FastICA extension}
\label{sec:3}
In the instantaneous case, \emph{FastICA} looks for a sequence of orthogonal projections to maximize the negentropy  $J(y_{i})$ \cite{AHy1998}, which amounts to seek components as independent as possible.
\begin{equation}
\label{neg}
\begin{aligned}
J(y_{i})&=\boldmath{H}(z_{i})-\boldmath{H}(y_{i})\\
&\approx c\left[\boldmath{E}\{G(y_{i}) \}-\boldmath{E}\{G(z_{i})\}\right]^{2}
\end{aligned}
\end{equation}
where  $J(y_{i})$ is the negentropy, $\boldmath{H}()$ is the random variable's entropy, $c$ is an irrelevant constant, $G()$ is any non-quadratic function, $z_{i}$ is a Gaussian random variable with the same variance as $y_{i}$. Mututal information $\boldmath{I}(\mathbf{y})$ between the components of the random variable vector $\mathbf{y}(k)$ is a nature measure of dependence. It is always non-negative and becomes zero only when the components are statistically independent. 
\begin{equation}
\label{info}
\begin{aligned}
\boldmath{I}(\mathbf{y})&=\sum_{i=1}^{m}\boldmath{H}(y_{i})-\boldmath{H}(\mathbf{y})\\
&=\sum_{i=1}^{m}\boldmath{H}(y_{i})-\boldmath{H}(\mathbf{x})-log\left|det(\mathbf{B})\right|\\
&=C-\sum_{i=1}^{m}J(y_{i})
\end{aligned}
\end{equation}
where $C$ is an irrelevant constant, the minimization of the mutual information $\boldmath{I}(\mathbf{y})$
for independence (under the constraint of decorrelation) is equivalent to the maximization of the sum of the negentropies of the components $\sum_{i=1}^{m}J(y_{i})$, and the original FastICA in instantaneous can be modeled as the optimization problem\cite{AHy1999} below,
\begin{equation}
\label{insopt}
\begin{aligned}
\max \quad & \sum_{i=1}^{m}J(y_{i}) & \\
\mathrm{s.t.} \quad & \boldmath{E}\{y_{i}(k)y_{j}(k)\} =\delta_{ij}, &\quad i,j=1,2,\ldots,m \,.
\end{aligned}
\end{equation}
where $\delta_{ij}$ is the item in identity matrix $\mathbf{I}_{m\times m}$. The objective function in (\ref{insopt}) aims at the maximization of independence. In both the deflation and symmetric mode, the optimization problem in (\ref{insopt}) is divided as single maximization of $J(y_{i})$, and there are 2$m$  extreme points to this problem due to the ordering and scaling ambiguities in instantaneous case. The constraint in (\ref{insopt}) is designed to avoiding extracting the same solution more than once.

While in the convolutive case, the filtering ambiguity introduces much more extreme points as the increasing of the demixing filters $\mathbf{B}$'s filter order $Q$, resulting in the failure of the same strategies in instantaneous BSS. It's straightforward to change the equation (\ref{insopt}) in the convolutive context,
\begin{equation}
\label{conopt}
\begin{aligned}
\max \quad & \sum_{i=1}^{m}J(y_{i}) & \\
\mathrm{s.t.} \quad & \boldmath{E}\{y_{i}(k)y_{i}(k)\} =1, &\quad i=1,2,\ldots,m \,,\\
& \boldmath{E}\{y_{i}(k)y_{j}(k-l)\} =0, &\quad i\neq j,-\infty<l<+\infty \,.
\end{aligned}
\end{equation}
The first constraint in (\ref{conopt}) is for the scaling ambiguity, and the second constraint is designed to tack the ordering and filtering ambiguities. In order to simplify the second constraint, we choose $-L\leq l\leq L$, where $L$ is a positive large enough integer. Prewhitening stage and diagonalization constraints in our modifications are used to satisfy these constraints.
\section{Prewhitening and diagonalization constraints}
\label{sec:4}
Many contrast function methods for convolutive mixtures have a prewhiteing stage (\ref{prew}) in space and time\cite{handbook2010}\cite{scd2007}.
\begin{equation}
\boldmath{E}\{\mathbf{v}(k)\mathbf{v}^{T}(k-l)\}=\delta_{l}\mathbf{I}_{n\times n} \quad \forall l \in Z
\end{equation}
The whitening filter $\mathbf{H}$ is not unique and hard to determine, the contrast functions are required to be  optimized under the constraint of complicate para-unitary filters $\mathbf{W}$. Although several methods have been proposed to alleviated these difficulties\cite{pc2003}\cite{jG2007}, they are both difficult and demands mass computing. We conduct a same prewhitening stage in instantaneous case\cite{dye1994} on $\underline {\mathbf{x}}(k)$ to produce the $nQ\times1$ column vector $\mathbf{v}(k)$, which is regarded as the convolutive prewhitening.
\begin{equation}
\label{precov}
\boldmath{E}\{\mathbf{v}(k)\mathbf{v}^{T}(k)\}=\mathbf{I}_{nQ\times nQ}	
\end{equation}
After the above prewhitening stage, the main effort is to adjust the seperating filters $\mathbf{W}$ to optimize equation (\ref{conopt}),  
\begin{equation}
\label{conopt2}
\begin{aligned}
\max \quad & \sum_{i=1}^{m}J(\mathbf{w}^{T}_{i}\mathbf{v}(k)) & \\
\mathrm{s.t.} \quad & \mathbf{w}^{T}_{i}\mathbf{w}_{i} =1, &\quad i=1,2,\ldots,m \,,\\
& \mathbf{w}^{T}_{i}\boldmath{E}\{\mathbf{v}^{T}(k)\mathbf{v}^{T}(k-l)\}\mathbf{w}_{j} =0, &\quad i\neq j,-L\leq l\leq L \,.
\end{aligned}
\end{equation}
where $\mathbf{w}^{T}_{i}$ is the $i$th row of $m\times nQ$ matrix $\mathbf{W}$. In pursuit of particular $\mathbf{w}_{i}$, the following iteration routine is carried out until convergence. 
\begin{equation}
\label{fix}
\begin{aligned}
\mathbf{w}_{i}&=\boldmath{E}\{\mathbf{v}(k)g(\mathbf{w}^{T}_{i}\mathbf{v}(k))\}-\boldmath{E}\{g^{'}(\mathbf{w}^{T}_{i}\mathbf{v}(k))\}\mathbf{w}_{i}\\
\mathbf{w}_{i}&=\mathbf{w}_{i}/\|\mathbf{w}_{i}\|_{2}
\end{aligned}
\end{equation}
where $g()$ is the derivative of particular non-quadratic function $G()$, coefficients constraints are required during the process of (\ref{fix}) in deflation and symmetric mode.
\par The validity of constraint in (\ref{conopt})(\ref{conopt2}) can also supported from \cite{dye1994}\cite{KAWAMOTO1998157} the fact: If the source signals have unique temporal structures or non-stationary, simultaneous diagonalization of ouput correlation matrices over multiple time lags can separate the independent sources from convolutive mixtures.
\begin{equation}
\label{Ry}
\begin{aligned}
\mathbf{R}_{y}(\tau)&=\boldmath{E}\{\mathbf{y}(k)\mathbf{y}^{T}(k-\tau)\}\\
&=\mathbf{W}\boldmath{E}\{\mathbf{v}(k)\mathbf{v}^{T}(k-\tau)\}\mathbf{W}^{T}\\
&=\mathbf{W}\mathbf{R}_{v}(\tau)\mathbf{W}^{T}
\end{aligned}
\end{equation}
The output correlation matrix at time lag $\tau$ is represented as $\mathbf{R}_{y}(\tau)$, it is required to be a diagonal matrix, particularly,$\mathbf{R}_{y}(0)$ is the identity matrix $\mathbf{I}_{m\times m}$. For more intuitive explanation, we define several column vectors.
\begin{equation}
\label{y1}
\underline {\mathbf{y}}(k)=\left(\mathbf{y}_{1}^{T}(k),\mathbf{y}_{2}^{T}(k),\cdots,\mathbf{y}_{m}^{T}(k)\right)^{T}
\end{equation}
\begin{equation}
\label{y2}
\mathbf{y}_{i}(k) = \left(y_{i}(k+L),y_{i}(k+L-1),\cdots,y_{i}(k-L)\right)^{T}
\end{equation}
where $\underline{\mathbf{y}}(k)$ is a $m(2L+1)\times1$ column vector, and $\mathbf{y}_{i}(k)$ is a $(2L+1)\times1$ column vector, the equivalent expression in (\ref{Ry}) can be described below,
\begin{equation}
\label{Ry1}
\begin{aligned}
\mathbf{R}_{\underline{\mathbf{y}}}&=\boldmath{E}\{\underline{\mathbf{y}}(k)\underline{\mathbf{y}}^{T}(k)\}\\
&=\left(\begin{IEEEeqnarraybox*}[][c]{,c/c/c/c,}
\mathbf{R}_{\mathbf{y}_{1}\mathbf{y}_{1}}&\mathbf{R}_{\mathbf{y}_{1}\mathbf{y}_{2}}&\cdots &\mathbf{R}_{\mathbf{y}_{1}\mathbf{y}_{m}}\\
\mathbf{R}_{\mathbf{y}_{2}\mathbf{y}_{1}}&\mathbf{R}_{\mathbf{y}_{2}\mathbf{y}_{2}}&\cdots &\mathbf{R}_{\mathbf{y}_{2}\mathbf{y}_{m}}\\
\vdots&\vdots &\ddots &\vdots\\
\mathbf{R}_{\mathbf{y}_{m}\mathbf{y}_{1}}&\mathbf{R}_{\mathbf{y}_{m}\mathbf{y}_{2}}&\dots &\mathbf{R}_{\mathbf{y}_{m}\mathbf{y}_{m}}%
\end{IEEEeqnarraybox*}\right)
\end{aligned}
\end{equation}
\begin{equation}
\label{Ry2}
\mathbf{R}_{\mathbf{y}_{i}\mathbf{y}_{j}}=\boldmath{E}\{\mathbf{y}_{i}(k)\mathbf{y}^{T}_{j}(k)\}
\end{equation}
where $m(2L+1)\times m(2L+1)$ matrix $\mathbf{R}_{\underline{\mathbf{y}}}$ is the combination of correlation matrix concerning different time lags. According to the constraint in (\ref{conopt}) and the nonstationarity property of source signals, the $(2L+1)\times(2L+1)$ matrix $\mathbf{R}_{\mathbf{y}_{i}\mathbf{y}_{j}}$ becomes nonzero only in the diagonal position of $\mathbf{R}_{\underline{\mathbf{y}}}$. We consider the coefficents constraints above as diagonalization constraints.
\par Compared with the existing para-unitary constraint, the diagonalization constraints are relaxed, it can be efficiently imposed via singular value decomposition in both deflation and symmetric mode.
\subsection{deflation mode}
\par When one of the sources' estimation $y_{i}(k)$ has been extracted, it's necessary to substract its contribution from the observations to obtain the mixtures of $m-1$ sources, then we repeat this procedure to extract the remained sources one by one, until all the sources have been extracted.
\par Considering the constraints in (\ref{conopt2}), we construct the block matrix $\mathbf{O}$ during the extraction of the $y_{i}(k)$ to guarantee uniqueness.
\begin{equation}
\label{block}
\mathbf{O} =\left(\mathbf{R}_{v}(-L)\mathbf{w}_{1},\mathbf{R}_{v}(-L+1)\mathbf{w}_{1},\cdots,\mathbf{R}_{v}(L)\mathbf{w}_{i-1}\right)
\end{equation}
\begin{equation}
\label{otrh}
\mathbf{w}^{T}_{i}\mathbf{O}=\mathbf{0}
\end{equation}
\par The $\mathbf{w}_{i}$ in the $i$th extraction is required to be orthogonal to the column space of the  block matrix $\mathbf{O}$, we represent the column space of $\mathbf{O}$ as $\boldmath{span}\{\mathbf{O}\}$.
\par If the block matrix $\mathbf{O}$ is overdetermined,the column space $\boldmath{span}\{\mathbf{O}\}$ is not full column rank, it's easy adjust the $\mathbf{w}_{i}$ via least square solution.
\begin{equation}
\label{lss}
\mathbf{w}_{i} =\mathbf{w}_{i} -\left(\mathbf{O}^{T}\mathbf{O}\right)^{-1}\mathbf{O}^{T}\mathbf{w}_{i}
\end{equation}
\par Unfortunately, $\mathbf{O}$ is often designed to be underdetermined due to choice of large $L$, so the strategy in (\ref{lss}) cannot work any longer. The proposed method draws lessons from the principal component analysis based on singular value decomposition, only taking the directions with most variation in $\mathbf{O}$ into consideration.
\begin{equation}
\label{svd}
\mathbf{O} =\mathbf{U}\mathbf{\Sigma}\mathbf{V}^{T}
\end{equation}
Here $\mathbf{U}$ and $\mathbf{V}$ are orthogonal matrices, with the columns of $\mathbf{U}$ spanning the column space of $\mathbf{O}$, and the columns of $\mathbf{V}$ spanning the row space. $\mathbf{\Sigma}$ is a diagonal matrix, with diagonal entries in decreasing order.
\begin{equation}
\sigma_{11}\ge\sigma_{22}\ge\cdots\ge0
\end{equation}
The algorithm picks the first $r$ columns vectors in $\mathbf{U}$ to obtain the most variance in original $\boldmath{span}\{\mathbf{O}\}$ based on the effective rank of $\mathbf{O}$, $r$ is the minimum integer when $\mu(r)$ is greater than particular threshold $\alpha$ (such as $0.99995$).
\begin{equation}
\label{effrank}
\mu(r) =\frac{\sqrt{\sigma_{11}^{2}+\cdots+\sigma_{rr}^{2}}}{\|\Sigma\|_{F}}
\end{equation}
The most variant $r$ column vectors in  $\mathbf{U}$ is represented as $\mathbf{U}_{(r)}$, and each column vector in $\mathbf{U}_{(r)}$ is orthogonal with each other, $\mathbf{U}_{(r)}$ is always high matrix,the (\ref{lss}) can be described in simple form.
\begin{equation}
\label{def1}
\mathbf{w}_{i} =\mathbf{w}_{i} -\mathbf{U}_{(r)}\mathbf{U}_{(r)}^{T}\mathbf{w}_{i}
\end{equation}
\par The  FastICA convolutive algorithm in deflation mode is summarized in Alg.$\ref{alg:def}$, where $tol$ is the threshold in iteration (such as $10^{-7}$).

\begin{algorithm}[!t]  
	\caption{convolutive FastICA: deflation mode }  
	\label{alg:def}  
	\begin{algorithmic}  
		\STATE {$\mathbf{W}=\mathbf{0}_{m\times nQ}$}
		\STATE {$i=1$}   
		\REPEAT   
		\STATE {$\mathbf{w}_{i}=\mathbf{0}_{nQ\times 1}$}
		\STATE{$\mathbf{O} =\left(\mathbf{R}_{v}(-L)\mathbf{w}_{1},\mathbf{R}_{v}(-L+1)\mathbf{w}_{1},\cdots,\mathbf{R}_{v}(L)\mathbf{w}_{i-1}\right)$}
		\STATE{calculate $\mathbf{U}_{(r)}$ from (\ref{svd})(\ref{effrank}) with threshold $\alpha$} 
		\REPEAT
		\STATE {$\mathbf{w}^{'}_{i}=\mathbf{w}_{i}$}
		\STATE{$\mathbf{w}_{i}=\boldmath{E}\{\mathbf{v}(k)g(\mathbf{w}^{T}_{i}\mathbf{v}(k))\}-\boldmath{E}\{g^{'}(\mathbf{w}^{T}_{i}\mathbf{v}(k))\}\mathbf{w}_{i}$}
		\STATE{$\mathbf{w}_{i} =\mathbf{w}_{i} -\mathbf{U}_{(r)}\mathbf{U}_{(r)}^{T}\mathbf{w}_{i}$}
		\STATE{$\mathbf{w}_{i}=\mathbf{w}_{i}/\|\mathbf{w}_{i}\|_{2}$}
		\UNTIL{$||\mathbf{w}^{T}_{i}\mathbf{w}^{'}_{i}|-1.0|\leq tol$}
		\STATE {$\mathbf{W}\left[i,:\right]=\mathbf{w}_{i}$}  
		\STATE {$i=i+1$}   
		\UNTIL{$i>m$}  
	\end{algorithmic}  
\end{algorithm}

\subsection{symmetric mode}
\par Every extracted source has different priority in deflation mode, leading an accumulation of estimation errors. Symmetric mode is proposed to extract $\mathbf{w}_{i}$ equally. Supposing only the $\mathbf{w}_{i}$ is in process, other extracted sources are fixed. It's nature to construct the block matrix from (\ref{conopt2})(\ref{block}).
\begin{equation}
\label{block2}
\mathbf{O} =\left(\mathbf{R}_{v}(-L)\mathbf{w}_{1},\cdots,\mathbf{R}_{v}(L)\mathbf{w}_{i-1},\mathbf{R}_{v}(-L)\mathbf{w}_{i+1},\cdots\right)
\end{equation}
The FastICA convolutive algorithm in symmetric mode is similar to the deflation mode, and it can be summarized in Alg.\ref{alg:sym}.
\begin{algorithm}[!t]  
	\caption{convolutive FastICA: symmetric mode }  
	\label{alg:sym}  
	\begin{algorithmic}  
		\STATE {$\mathbf{W}=\mathbf{0}_{m\times nQ}$}
		\REPEAT
		\STATE {$\mathbf{W}^{'}=\mathbf{W}$}
		\STATE {$i=1$}
		\REPEAT
		\STATE {$\mathbf{w}_{i}=\mathbf{W}\left[i,:\right]$}     
		\STATE{$\mathbf{O} =\left(\mathbf{R}_{v}(-L)\mathbf{w}_{1},\cdots,\mathbf{R}_{v}(L)\mathbf{w}_{i-1},\mathbf{R}_{v}(-L)\mathbf{w}_{i+1},\cdots\right)$}
		\STATE{calculate $\mathbf{U}_{(r)}$ from (\ref{svd})(\ref{effrank}) with threshold $\alpha$} 
		\STATE{$\mathbf{w}_{i}=\boldmath{E}\{\mathbf{v}(k)g(\mathbf{w}^{T}_{i}\mathbf{v}(k))\}-\boldmath{E}\{g^{'}(\mathbf{w}^{T}_{i}\mathbf{v}(k))\}\mathbf{w}_{i}$}
		\STATE{$\mathbf{w}_{i} =\mathbf{w}_{i} -\mathbf{U}_{(r)}\mathbf{U}_{(r)}^{T}\mathbf{w}_{i}$}
		\STATE{$\mathbf{w}_{i}=\mathbf{w}_{i}/\|\mathbf{w}_{i}\|_{2}$}
		\STATE {$\mathbf{W}\left[i,:\right]=\mathbf{w}_{i}$}  
		\STATE {$i=i+1$}   
		\UNTIL{$i>m$}
		\UNTIL{$\||\mathbf{W}^{'}\mathbf{W}^{T}|-\boldmath{I}_{m\times m}\|_{2}\leq tol$}  
	\end{algorithmic}  
\end{algorithm}  

\subsection{sources' reconstruction}
\par After the extraction of $\mathbf{y}(k)$ in deflation or symmetric mode, we get the  estimation of the ordered, scaled, and delayed version of innovation process $\mathbf{u}(k)$ (\ref{innovation}), we then conduct a reconstruction process to recover the signals $\mathbf{s}(k)$'s contributions in the observation mixtures $\mathbf{x}(k)$. In this part,for the convenience of discussing, the ordering ambiguity is ignored.
\par For the purpose of achieve innovation $s_{i}(k)$'s contributions in observation $x_{j}(k)$, a $(2L+N)\times(2L+1)$ high matrix $\mathbf{T}$ is built to represent the column space via zero padding, including all the possible forward and backward time shift of $y_{i}$.
\begin{equation}
\label{uspace}
\mathbf{T} =\left(\begin{IEEEeqnarraybox*}[][c]{,c/c/c/c/c,}
y_{i}(0)&0&\cdots &0&0\\
y_{i}(1)&y_{i}(0)&\cdots &0&0\\
y_{i}(2)&y_{i}(1)&\cdots &0&0\\
\vdots&\vdots &\ddots &\vdots&\vdots\\
0&0&\cdots &y_{i}(N-2)&y_{i}(N-3)\\
0&0&\cdots &y_{i}(N-1)&y_{i}(N-2)\\
0&0&\cdots &0&y_{i}(N-1)%
\end{IEEEeqnarraybox*}\right)
\end{equation}
$N\times 1$ column vector $x_{j}$ is padding with $L$ zeros both forward and backward.
\begin{equation}
\label{xj2}
x_{j}^{(2L+N)}=\left(0,0,\cdots,x_{j}(0),x_{j}(1),\cdots,0,0\right)^{T}
\end{equation}
The reconstruction is based on the regression opinion: regress $x_{j}^{(2L+N)}$ on the column space of $\mathbf{T}$, finding the closest $\hat{s}_{ij}$ in the $\boldmath{span}\{\mathbf{T}\}$ in least square sense. We describe the $s_{i}$ contribution to observation $x_{j}$ as $\hat{s}_{ij}$. 
\begin{equation}
\label{regress}
\hat{s}_{ij} =(\mathbf{T}^{T}\mathbf{T})^{-1}\mathbf{T}^{T}x^{(2L+N)}_{j}
\end{equation}
\section{Experiment results}
\label{sec:5}
We conducted two experiments in this section to explore the performance of the proposed algorithm.
\begin{figure}[!t]
	\centering
	\subfloat[Source signals]{\includegraphics[width=2.5in]{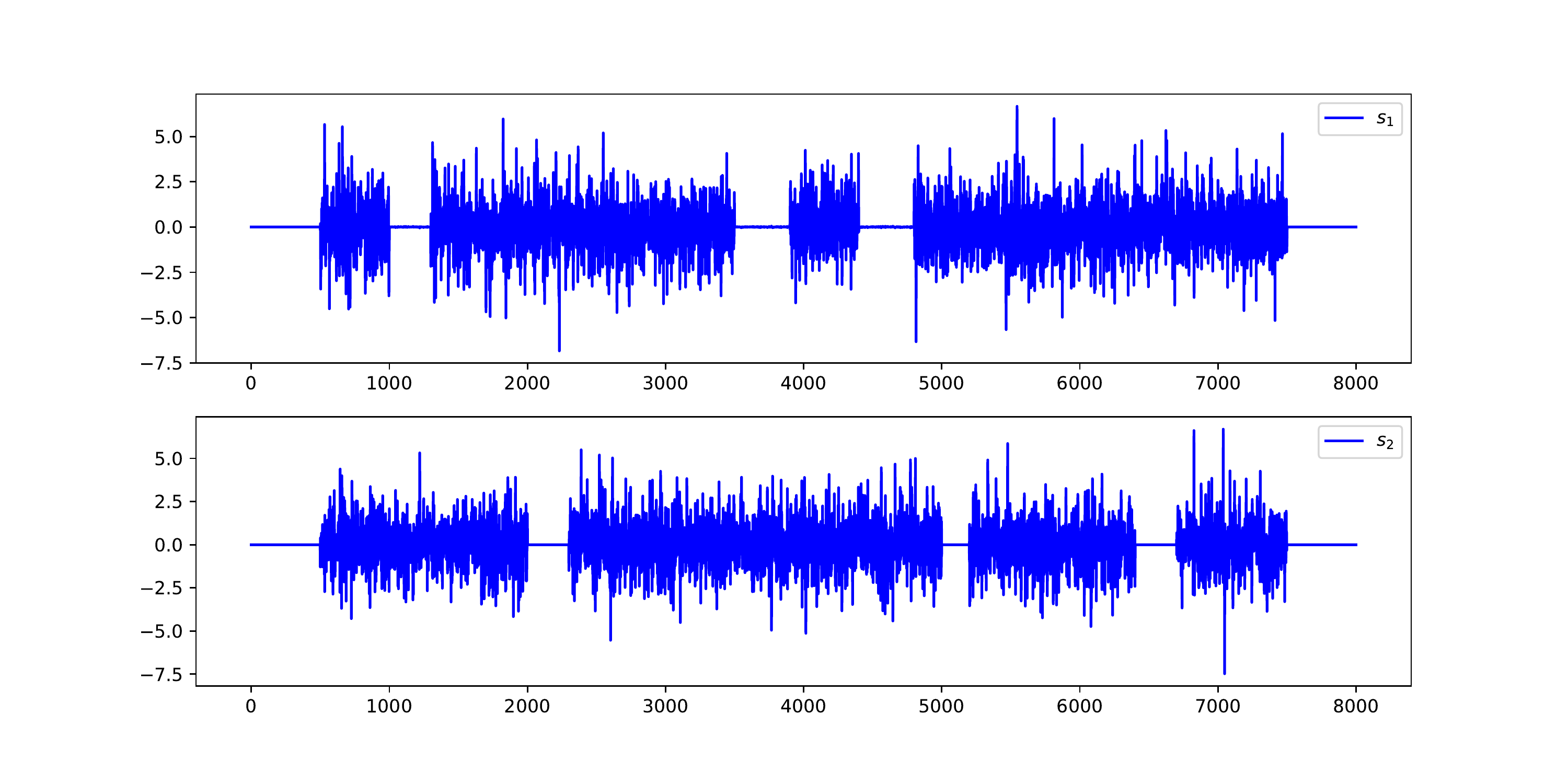}
		\label{2x2s}}
	\vfil
	\subfloat[Observations]{\includegraphics[width=2.5in]{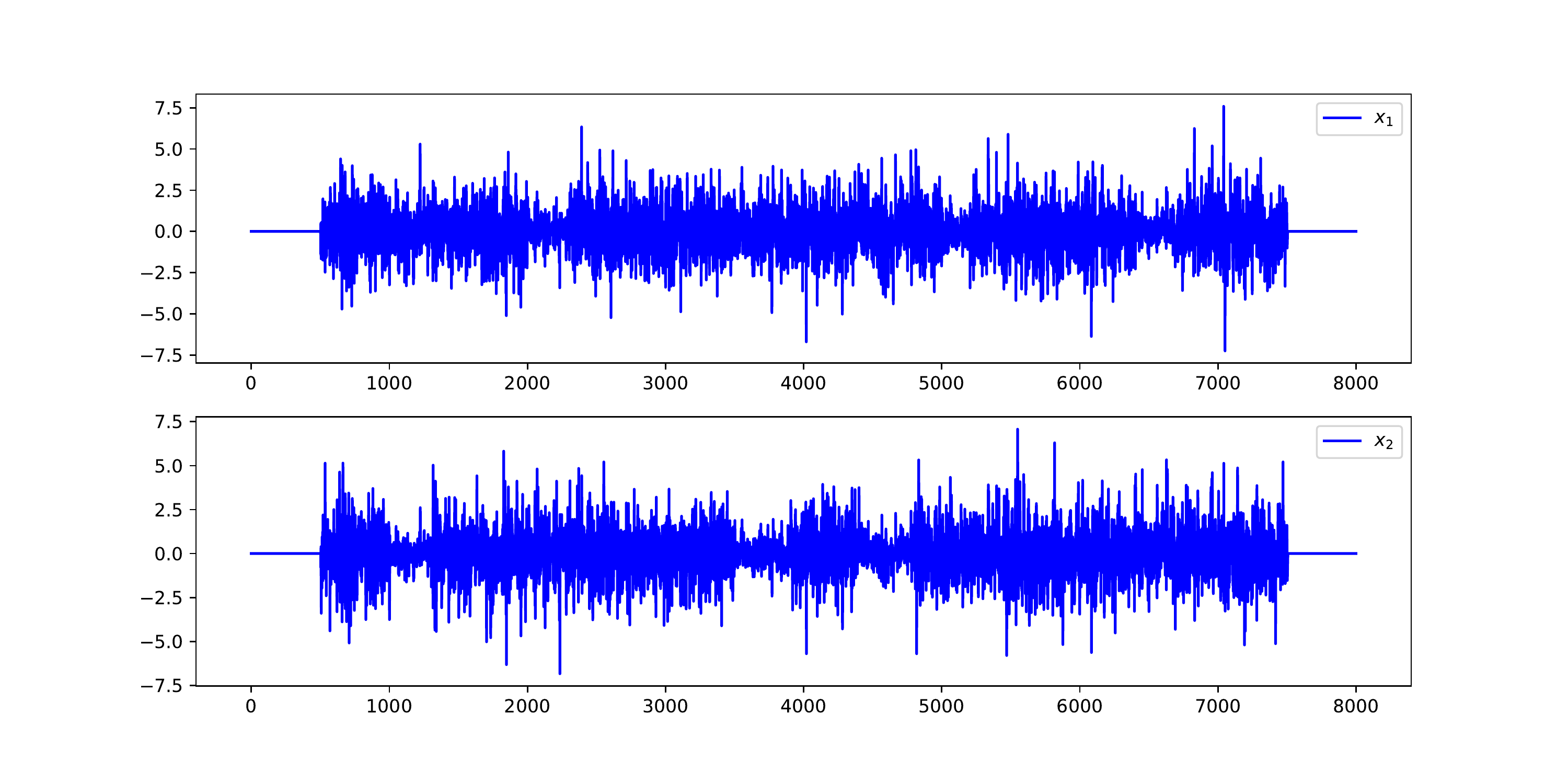}
		\label{2x2o}}
	\caption{Source signals and observations in the 2$\times$2 case(simulation).}
	\label{2x21}
\end{figure}
\par The first experiment mixed the given innovation process $u_{1}$ and $u_{2}$ in Fig.\ref{2x2s}, both satisfying the referred assumptions. After the convolutive prewhitening stage, the algorithm in symmetric mode recovered the $y_{1}$ and $y_{2}$, and calculated their contributions on each observation.
\begin{figure}[!t]
	\centering
	\subfloat[Innovation process]{\includegraphics[width=2.5in]{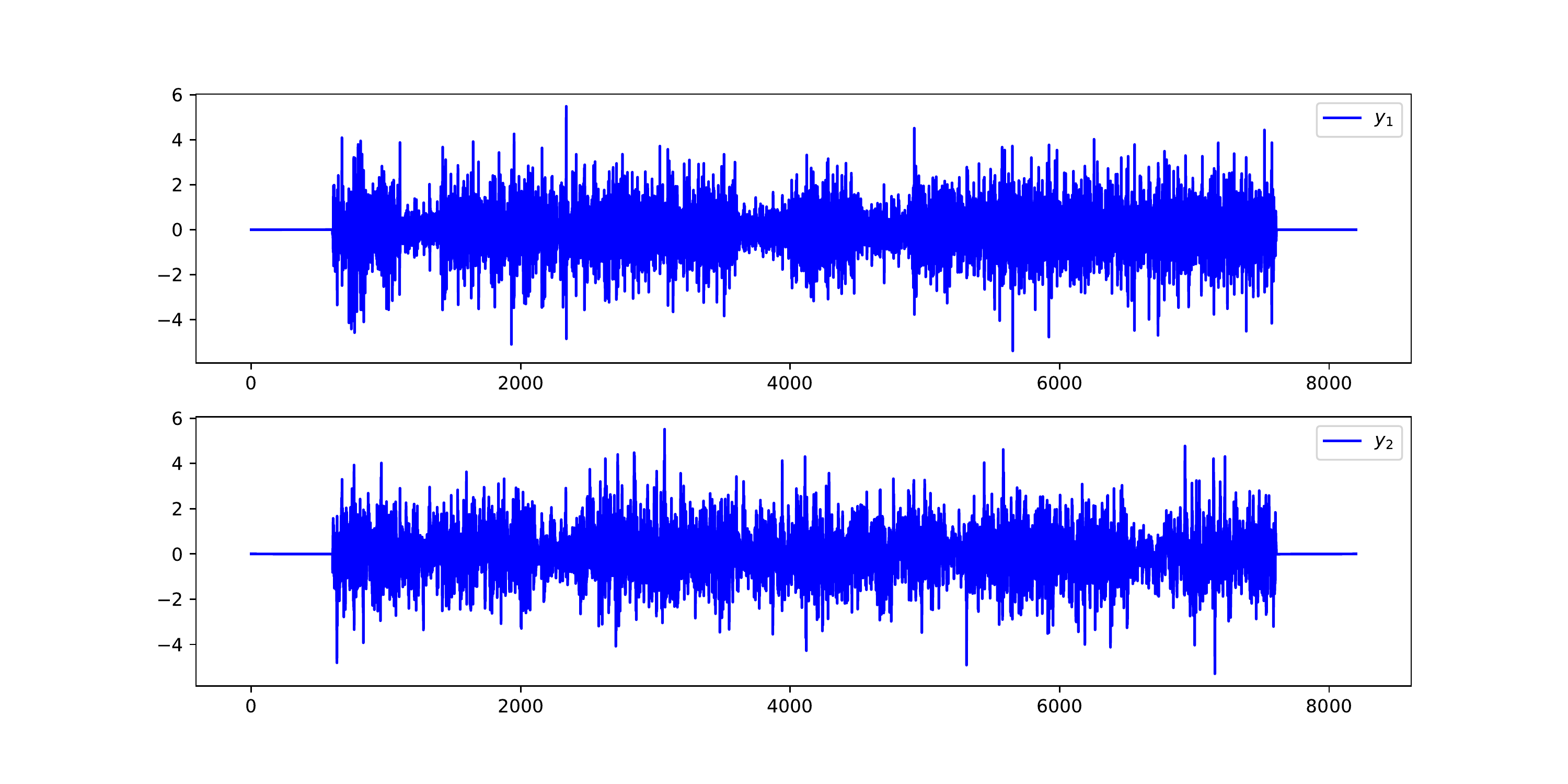}
		\label{2x2i}}
	\vfil
	\subfloat[Sources' contirbutions]{\includegraphics[width=2.5in]{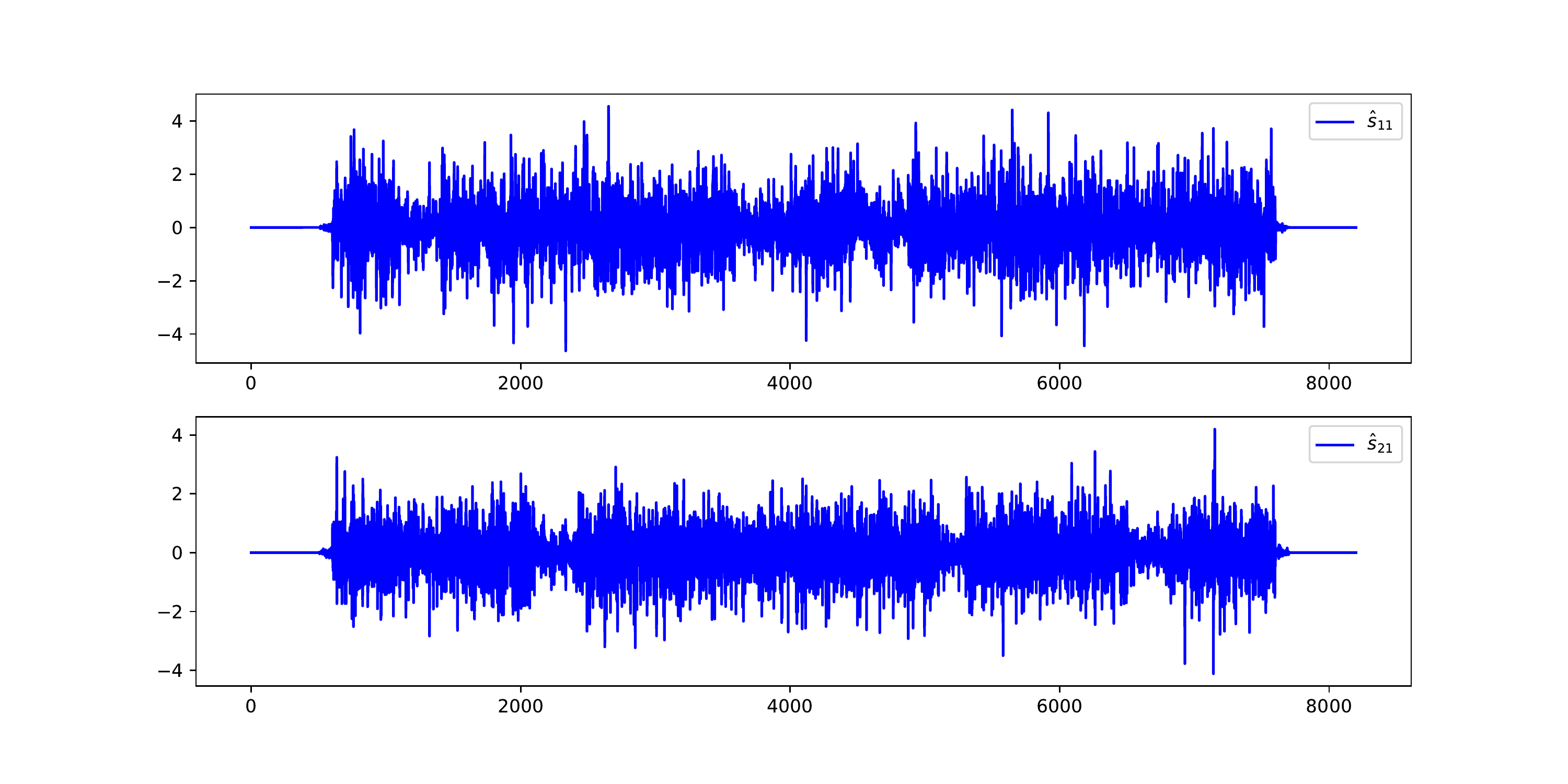}
		\label{2x2rs}}
	\caption{Innovation process and sources' contributions in the 2$\times$2 case(symmetric mode).}
	\label{2x22}
\end{figure}
\par In Fig.\ref{2x2i} the recovered estimation $y_{1}$ and $y_{2}$ are similar to the original innovation process, while there are little deviations in the orignal zero amplitude regions. The contributions in Fig.\ref{2x2rs} happened to be the same as innovation process. In this simulation.the algorithm accomplished the mission of convolutive BSS.
\par A more difficult task was considered in real recorded source signals from the public data of Salk Institute\cite{publicdata}.
\begin{enumerate}[\IEEEsetlabelwidth{12)}]
	\item $s_{1}$ :speaker says the digits from one to ten in English. 
	\item $s_{2}$ :loud music in the background.
\end{enumerate}

\begin{figure}[!t]
	\centering
	\subfloat[Source signals]{\includegraphics[width=2.5in]{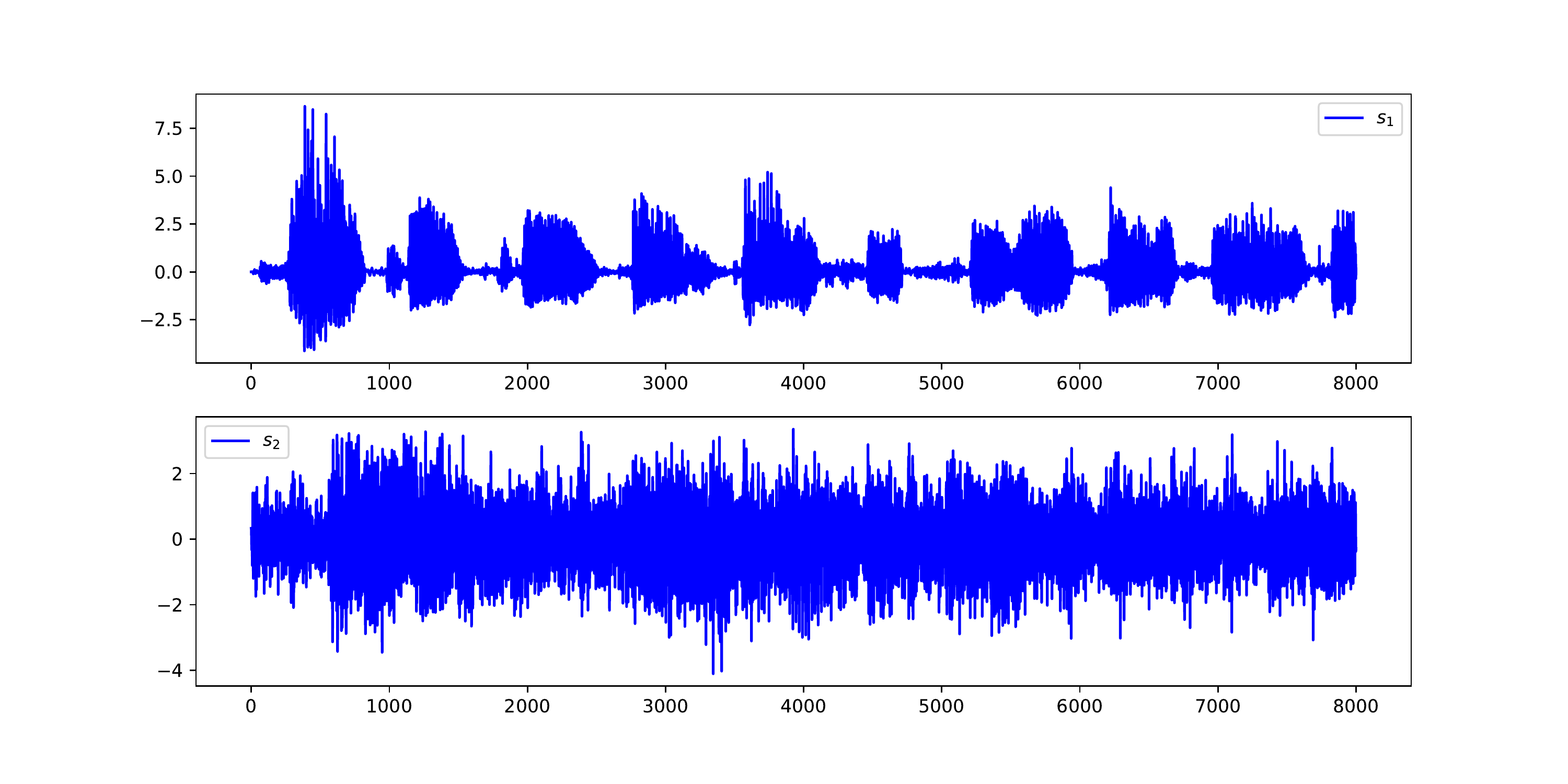}
		\label{3x3s}}
	\vfil
	\subfloat[Observations]{\includegraphics[width=2.5in]{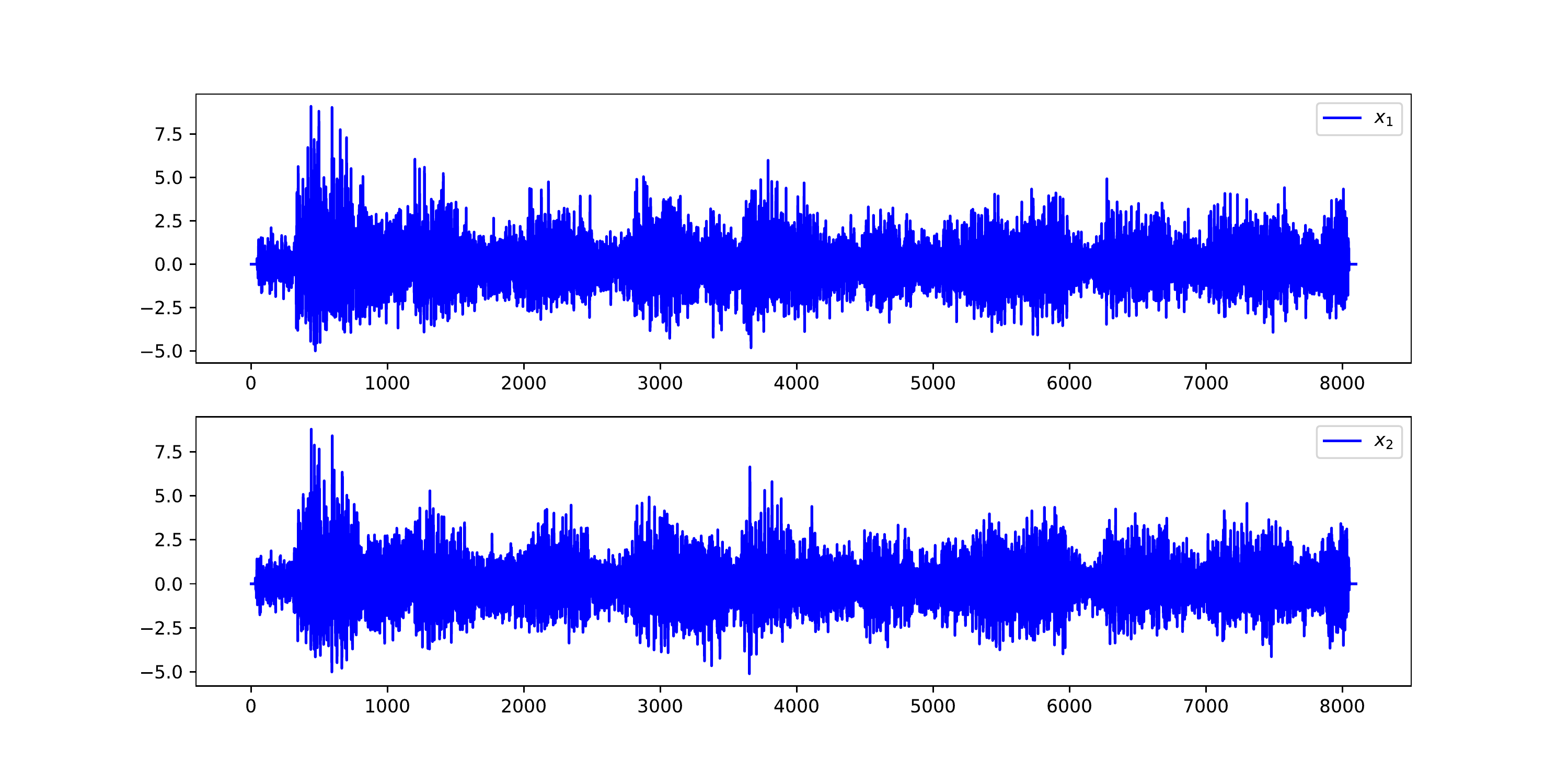}
		\label{3x3o}}
	\caption{Source signals and observations in the 2$\times$2 case(record).}
	\label{3x31}
\end{figure}
\par
\begin{figure}[!t]
	\centering
	\subfloat[Innovation process]{\includegraphics[width=2.5in]{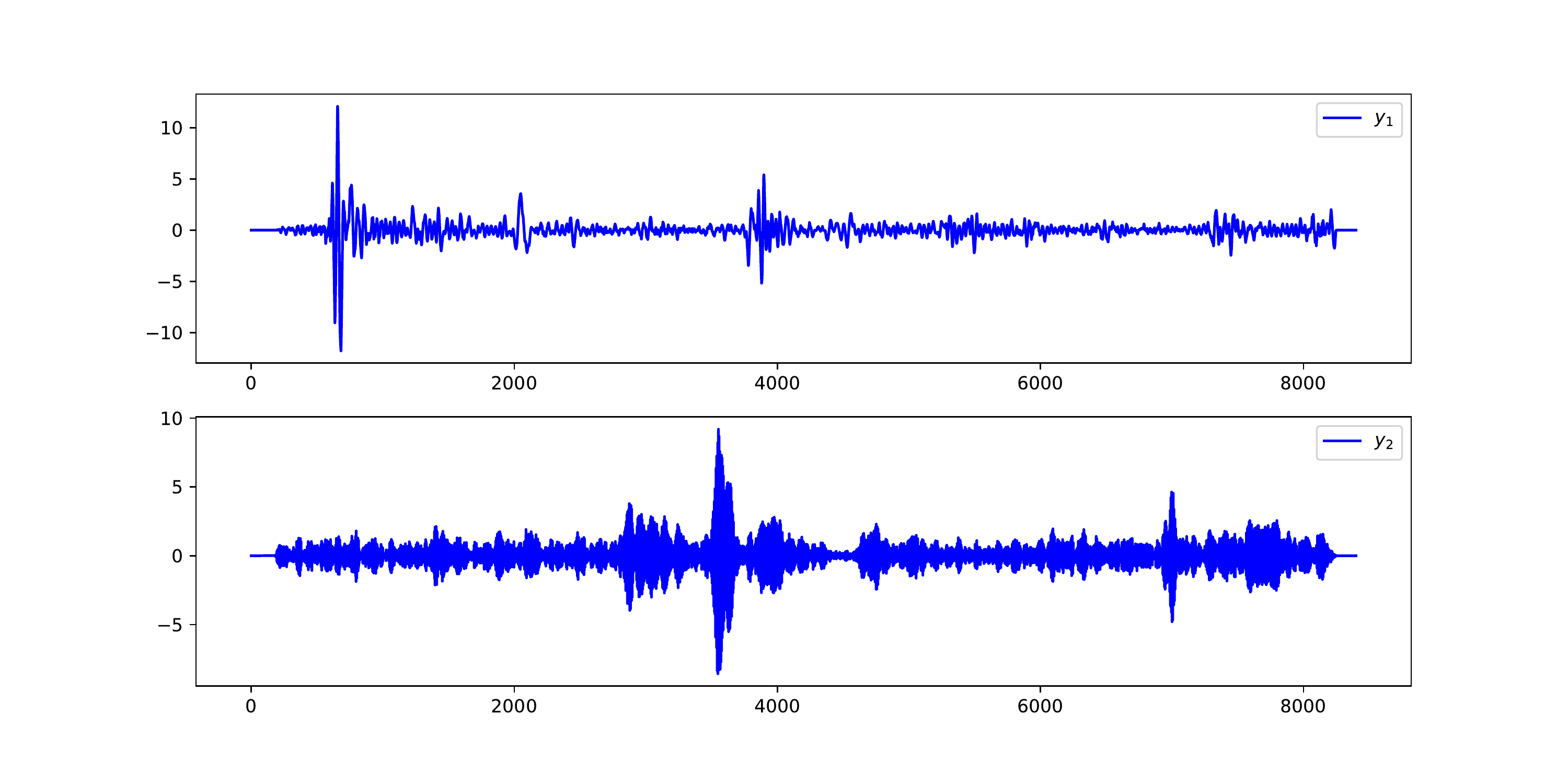}
		\label{3x3i}}
	\vfil
	\subfloat[Sources' contirbutions]{\includegraphics[width=2.5in]{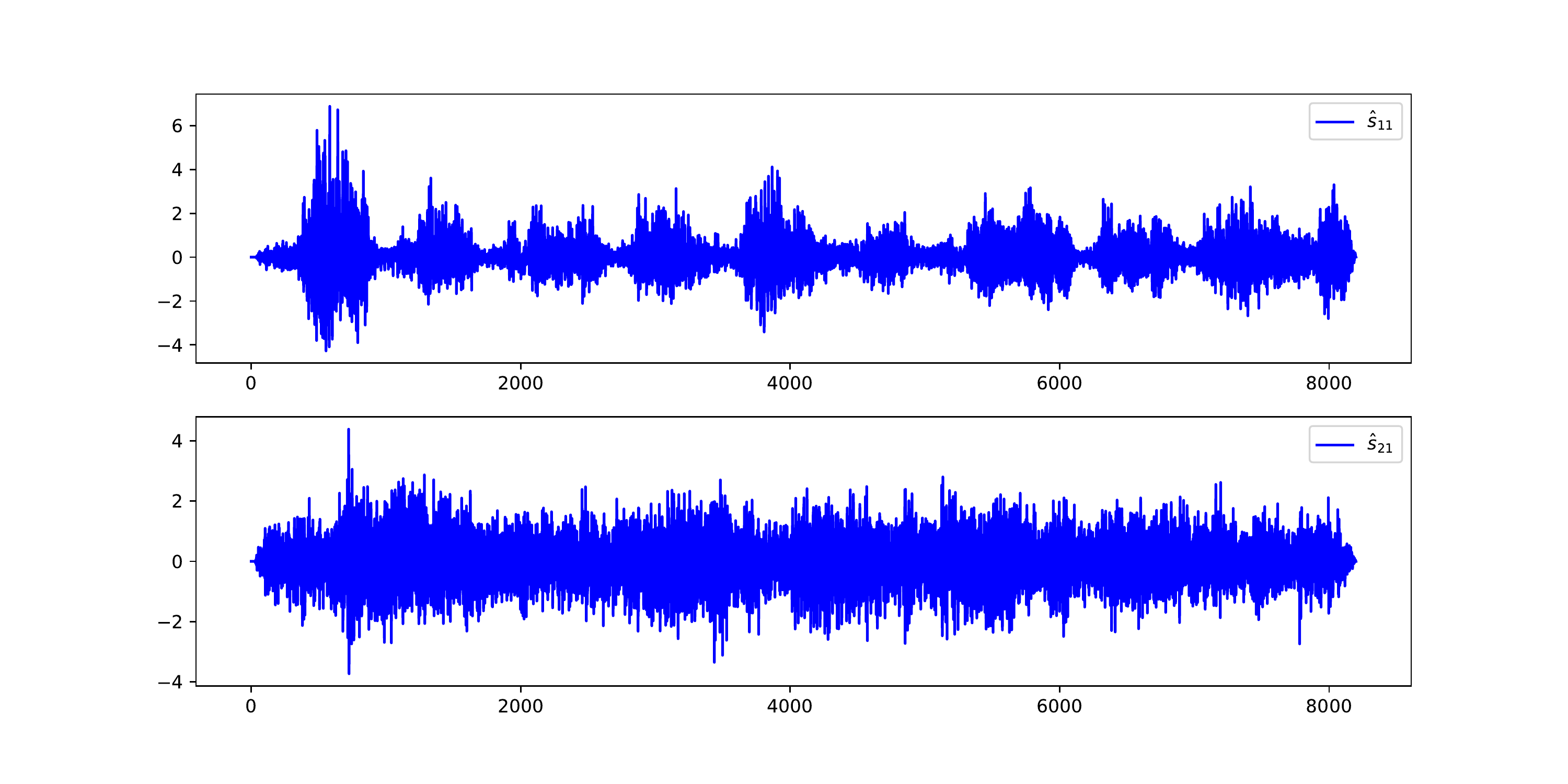}
		\label{3x3rs}}
	\caption{Innovation process and sources' contributions in the 2$\times$2 case(deflation mode).}
	\label{3x32}
\end{figure}

 We conduct deflation mode in Alg.\ref{alg:def} to produce the estimations and contributions. The results in Fig.\ref{3x3rs} showed the similarity between the calculated contributions and original source signals.

\section{Conclusion}
\label{sec:6}
\par In this paper, we have derived a novel extension of the \emph{FastICA} for convolutive mixtures that enforces the diagnoalization constraints on the seperating filters for uniqueness. Our algorithm also has simple convolutive prewhitening stages and contributions' reconstruction procedure. Experiments are given to illustrate the performance of the proposed algorithm.
\par Our algorithm enjoys the robustness and fast convergence due to its fix-point iterations, no particular parameter tuning is required during the optimization. Compared with spatial-temporal prewhitening stages and para-unitary filter constraints in other contrast function methods\cite{handbook2010}, the corresponding procedures in our algorithm are much more straightforward and simpler.

% if have a single appendix:
%\appendix[Proof of the Zonklar Equations]
% or
%\appendix  % for no appendix heading
% do not use \section anymore after \appendix, only \section*
% is possibly needed

% use appendices with more than one appendix
% then use \section to start each appendix
% you must declare a \section before using any
% \subsection or using \label (\appendices by itself
% starts a section numbered zero.)
%

% use section* for acknowledgment
%\section*{Acknowledgment}

%The authors would like to thank...

%\appendices
%\section{Proof of the First Zonklar Equation}
%Appendix one text goes here.

% Can use something like this to put references on a page
% by themselves when using endfloat and the captionsoff option.
\ifCLASSOPTIONcaptionsoff
  \newpage
\fi

% trigger a \newpage just before the given reference
% number - used to balance the columns on the last page
% adjust value as needed - may need to be readjusted if
% the document is modified later
%\IEEEtriggeratref{8}
% The "triggered" command can be changed if desired:
%\IEEEtriggercmd{\enlargethispage{-5in}}

% references section

% can use a bibliography generated by BibTeX as a .bbl file
% BibTeX documentation can be easily obtained at:
% http://mirror.ctan.org/biblio/bibtex/contrib/doc/
% The IEEEtran BibTeX style support page is at:
% http://www.michaelshell.org/tex/ieeetran/bibtex/
%\bibliographystyle{IEEEtran}
% argument is your BibTeX string definitions and bibliography database(s)
%\bibliography{IEEEabrv,../bib/paper}
%
% <OR> manually copy in the resultant .bbl file
% set second argument of \begin to the number of references
% (used to reserve space for the reference number labels box)
%\begin{thebibliography}{1}

%\bibitem{IEEEhowto:kopka}
%H.~Kopka and P.~W. Daly, \emph{A Guide to \LaTeX}, 3rd~ed.\hskip 1em plus
%  0.5em minus 0.4em\relax Harlow, England: Addison-Wesley, 1999.

%\end{thebibliography}

\bibliographystyle{IEEEtran}
\bibliography{convolutiveBSS}

% biography section
% 
% If you have an EPS/PDF photo (graphicx package needed) extra braces are
% needed around the contents of the optional argument to biography to prevent
% the LaTeX parser from getting confused when it sees the complicated
% \includegraphics command within an optional argument. (You could create
% your own custom macro containing the \includegraphics command to make things
% simpler here.)
%\begin{IEEEbiography}[{\includegraphics[width=1in,height=1.25in,clip,keepaspectratio]{mshell}}]{Michael Shell}
% or if you just want to reserve a space for a photo:

%\begin{IEEEbiography}{Michael Shell}
%Biography text here.
%\end{IEEEbiography}

% if you will not have a photo at all:
%\begin{IEEEbiographynophoto}{John Doe}
%Biography text here.
%\end{IEEEbiographynophoto}

% insert where needed to balance the two columns on the last page with
% biographies
%\newpage

%\begin{IEEEbiographynophoto}{Jane Doe}
%Biography text here.
%\end{IEEEbiographynophoto}

% You can push biographies down or up by placing
% a \vfill before or after them. The appropriate
% use of \vfill depends on what kind of text is
% on the last page and whether or not the columns
% are being equalized.

%\vfill

% Can be used to pull up biographies so that the bottom of the last one
% is flush with the other column.
%\enlargethispage{-5in}

% that's all folks
\end{document}